\documentstyle[editedvolume]{crckapb} 

\input{psfig}

\newcommand{\beqa}{\begin{eqnarray*}}
\newcommand{\eeqa}{\end{eqnarray*}}
\newcommand{\teff}{${T_{\rm eff}}$}
\newcommand{\feh}{${\rm [Fe/H]}$}
\newcommand{\uvby}{${\it uvby}$}

\newcommand{\by}{b$-$y}
\newcommand{\al}{\rm et al.}

\begin{opening}
\title{Simultaneous solutions of stellar parameters \protect\\
       ({\teff}, {\feh}) from synthetic {\uvby} Str{\"o}mgren photo-\protect\\ 
       metry}
 
\author{E. Lastennet}
\institute{Queen Mary \& Westfield College, Astronomy Unit, \\
           Mile End Road, London E1 4NS, UK}
\author{T. Lejeune}
\author{P. Westera}
\author{R. Buser}  
\institute{Astronomisches Institut der Universit\"at Basel,\\
           Venusstr. 7, CH--4102 Binningen, Switzerland}
\end{opening}

\runningtitle{Simultaneous solution of ({\teff},{\feh})}

\begin{document}

\section{Introduction}
The comprehensive knowledge  of fundamental parameters of single stars
is  the basis of the  modelling of star clusters   and galaxies.  Most
fundamental  stellar parameters of  the  individual components in  SB2
eclipsing binaries are known  with very high accuracy.  Unfortunately,
while masses and radii are well  determined, the temperatures strongly
depend on  photometric calibrations.  In the present work, we  have used 
an empirically-calibrated  grid of   theoretical  stellar  spectra (BaSeL
models) for simultaneously deriving homogeneous effective temperatures
and metallicities  from observed data.
For this purpose, we have selected 20 binary systems (40  stars) for 
which we  have $uvby$  Str{\"o}mgren  photometry with  estimated errors
(see Lastennet {\al} 1998 for details). 

\section{An example of simultaneous ({\teff},{\feh}) determination}
To compute synthetic colours from the BaSeL  models, we need effective
temperature (\teff), surface gravity ($\log g$), and  metallicity
({\feh}). Consequently, given the observed colours (namely, b$-$y,
m$_1$,  and c$_1$), we are able  to derive {\teff},  $\log g$, and {\feh}
from a comparison with model colours.  As the surface gravities can be
derived very accurately from the masses and radii  of the stars in our
working  sample, only two  physical parameters have  to  be  derived
({\teff} and {\feh}). This has been done by minimizing the 
$\chi^2$-functional, defined as
\beqa 
 \chi^2 (T_{\rm eff}, [Fe/H])  = \sum_{i=1}^{n} \left[ \left(\frac{\rm
 colour(i)_{\rm   syn} -  colour(i)}{\sigma(\rm   colour(i))}\right)^2
 \right],
\eeqa 
where  $n$ is the  number   of comparison data, colour(1)=  (\by)$_0$,
colour(2)=  m$_0$,  and  colour(3) =  c$_0$.    The  best $\chi^2$  is
obtained when the synthetic colour, colour(i)$_{\rm syn}$, is equal to
the observed one. Finding  the central minimum value
$\chi^{2}_{\rm min}$, we   form  the $\chi^2$-grid in   the  ({\teff},
{\feh})-plane and compute  the boundaries corresponding to 1 $\sigma$,
2 $\sigma$, and 3 $\sigma$ respectively, as shown in Fig.1 for one of 
the 40 stars. 

\begin{figure}[!h]
\centerline{\psfig{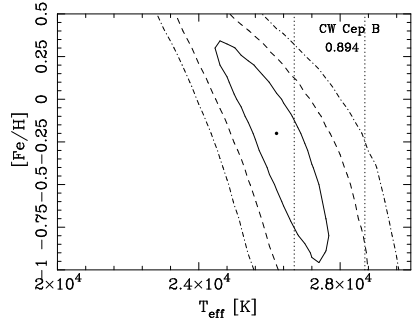}}
	\caption[]{Simultaneous solutions of {\teff}  and {\feh} for  CW
	Cep B. Best fit ({\it filled dot}) and 1-$\sigma$ ({\it
	solid line}), 2-$\sigma$ ({\it dashed line}), and
	3-$\sigma$({\it dot-dashed line}) confidence levels are also
	shown.  Previous estimates of {\teff} from Clausen \& Gim\'enez 
        (1991) are indicated as vertical dotted lines.}
\end{figure}

The range of {\teff} values that we derive  for CW Cep B agrees
well with previous  estimates (as indicated  by  the vertical dotted
lines),  and its metallicity is compatible with those of the Galactic disk stars.
As a general trend for the whole sample, it is worth noticing that our {\teff}  ranges do not provide estimates systematically  different from previous ones. 
However, the confidence regions show that most previous estimates of {\teff} 
are optimistic, and that {\feh} should not be neglected in such 
determinations. The  effects of surface  gravity and interstellar reddening 
have also been carefully studied. Moreover, comparisons for 8 binaries of 
our working sample with Hipparcos parallaxes show good agreement with the 
most reliable parallaxes. Full details and critical discussions are given 
in Lastennet {\al} (1998).

\end{document}